\documentclass[aps,nofootinbib,showpacs,amsmath,amssymb, preprint]{revtex4}
\usepackage{epsfig}
\usepackage{epsf}
\usepackage{amssymb}
\usepackage{amsmath}
\newcommand{\be}{\begin{equation}}
\newcommand{\ee}{\end{equation}}
\newcommand{\bea}{\begin{eqnarray}}
\newcommand{\eea}{\end{eqnarray}}
\begin{document}

\title{ On the ambiguity of field correlators represented \\ 
by asymptotic perturbation expansions} 
\author{Irinel Caprini} 
\affiliation{National Institute of Physics and Nuclear Engineering,  \\Bucharest POB MG-6, R-077125 Romania} 
\author{Jan Fischer}   
\affiliation{Institute of Physics, Academy of Sciences of the Czech Republic,
\\CZ-182 21  Prague 8, Czech Republic}  
\author{Ivo Vrko\v{c}}  
\affiliation{Mathematical Institute, Academy of Sciences of the Czech
Republic, 
\\CZ-115 67  Prague 1, Czech Republic}

\vskip1cm

\begin{abstract}

Starting from the divergence pattern of perturbation expansions in Quantum 
Field Theory and the (assumed) 
asymptotic character of the series, we address the problem of ambiguity of a  
function determined by the perturbation expansion. We consider functions 
represented by an  integral of the Laplace-Borel type along a general contour 
in the Borel complex plane. Proving a modified form of the Watson lemma, we 
obtain a large class of functions  having the same asymptotic perturbation 
expansion. Some remarks on perturbative QCD are made,  using the particular case of the Adler function.
\end{abstract}

%\keywords{perturbative QCD, divergent series} 
\maketitle

%%%%%%%%%%%%%%%%%%%%%%%%%%%%%%%%%%%%%%%%%%%%%%%%%%%%%%%%%%%%%%
\section{Introduction}
%%%%%%%%%%%%%%%%%%%%%%%%%%%%%%%%%%%%%%%%%%%%%%%%%%%%%%%%%%%%%%
It has been known for a long time that perturbation expansions in QED and QCD 
are, under plausible assumptions, divergent series. 
This result obtained by Freeman Dyson for QED \cite{Dyson} was a surprise in 
1952 and set a challenge for a radical reformulation of perturbation theory. 
Dyson's argument has been repeatedly critically revised and reformulated since
  \cite{Lautrup}-\cite{Mueller1985} (for a review see also \cite{Jan}), 
  with the conclusion that perturbation series appear to be 
divergent in many physically interesting situations. To give the divergent
series precise meaning, Dyson proposed to interpret it as asymptotic to $F(z)$,
the function searched for:
\begin{equation}
F(z) \,\, \sim  \,\, \sum_{n=0}^{\infty} F_{n} z^{n},\quad\quad\quad  
z \in {\cal S}, \quad  z \rightarrow 0,
\label{ptas}
\end{equation} 
where ${\cal S}$ is a region having the origin as an accumulation point, $z$ 
being the perturbation parameter. Dyson's assumption of asymptoticity has been 
widely adopted. 

By this, the philosophy of perturbation theory changed radically.  Perturbation 
theory yields, at least in principle, the values of all the $F_{n}$ coefficients. 
This can tell us whether the series is convergent or not, but what we want to 
know is under what conditions $F(z)$ can be determined from (\ref{ptas}).
If the series in (\ref{ptas}) were  convergent and the sign $\sim$ were replaced 
by  equality, the knowledge of all the $F_{n}$ would uniquely determine $F(z)$.   
On the other hand, there are infinitely many functions having the same {\it 
asymptotic} expansion (\ref{ptas}).

This situation raises the problem of finding the "correct" or "physical" 
function $F(z)$, using the knowledge of all the $F_{n}$ coefficients (or, in a 
more realistic situation, the knowledge of several first terms only) of the 
series. The infinite ambiguity of the solution of this problem may be reduced if 
a specific field theory or model is considered allowing one to exploit some 
additional inputs of the specific theory. Useful information can be found in the
papers \cite{KazSh}, \cite{KazPop} and the references therein.  

The objective of the present paper is to discuss the ambiguities of perturbation 
theory stemming from the assumed asymptotic character of the series. A class 
${\cal C}$ of functions admitting a given asymptotic expansion is specified by 
the lemma of Watson, which we recall in section \ref{sec:watlem}. Watson's lemma, 
on the other hand, does not imply that ${\cal C}$ is the maximal class of that 
kind. In section \ref{sec:modwat} we present, and in section \ref{sec:prooflem} 
we prove, a modified form of Watson's lemma. The modified lemma, which we refer 
to as Lemma 2 in this paper, allows  us to show that the class ${\cal C}'$ of 
functions possesing one given asymptotic expansion can be, under plausible 
conditions, much larger than ${\cal C}$. A discussion of Lemma 2 and its proof 
is placed in section \ref{sec:remlem}. We discuss some applications in section 
\ref{sec:remqcd}, using as an example the Adler function \cite{Adler} in QCD.

%%%%%%%%%%%%%%%%%%%%%%%%%%%%%%%%%%%%%%%%%%%%%%%%%%%%%%%%%%%%%%%%%%
\section{Perturbation theory and asymptotic series} \label{sec:ptas} 
%%%%%%%%%%%%%%%%%%%%%%%%%%%%%%%%%%%%%%%%%%%%%%%%%%%%%
\subsection{Perturbative approach}
%%%%%%%%%%%%%%%%%%%%%%%%%%%%%%%%%%%%
A typical difficulty in physics is lack of exact solutions. To find an 
approximation, one can neglect some effects, which can then be reintroduced as 
series in powers of some correction parameter, $z$, written generically as 
(\ref{ptas}), where $F(z)$ is the function searched for. It is assumed that the 
expansion coefficients $F_{n}$ are calculable from the theory. In most cases, 
however, only a few terms  have been calculated and, in QCD, we seem to be near 
the limit of what can be calculated within the available analytical and 
numerical tools. 

A typical question in the 1950's was whether a perturbation series of the type 
(\ref{ptas}) was convergent or not.  In many field theories and models, the 
large-order behavior of some subclasses of Feynman diagrams  shows that the series is divergent, the coefficients $F_n$ growing  as $n!$  
\cite{Dyson}-\cite{Jan}. But  a sum can, under certain conditions, be assigned 
even to a divergent series. So, the crucial problem is: does (\ref{ptas}) 
determine $F(z)$  uniquely, or not? The answer depends on additional inputs 
and, also, on how the symbol $\sim$ in (\ref{ptas}) is interpreted. 
%%%%%%%%%%%%%%%%%%%%%%%%%%%%%%%%%%%%%%%%%%%%%%%%%%%%%%%%%%%%%%%%%%%%%%
\subsection{Basic properties of asymptotic series} 
%%%%%%%%%%%%%%%%%%%%%%%%%%%%%%%%%%%%%%%%%
{\bf Definition:} Let ${\cal S}$ be a region or point set containing the 
origin or at least having it as an accumulation point. The power series 
$\sum_{n=0}^{\infty}F_{n}z^{n}$ is said to be asymptotic to the function $F(z)$ 
as $z \to 0$ on ${\cal S}$, and we write
Eq. (\ref{ptas}), if the set of functions $R_{N}(z)$,
\begin{equation}
R_{N}(z) = F(z) - \sum_{n=0}^{N}F_{n}z^{n} ,
\label{rema}
\end{equation}
satisfies the condition
\begin{equation}
R_{N}(z) = o(z^{N}) 
\label{ordo}
\end{equation}
for all $N=0,1,2,...$, $z \rightarrow 0$ and $z \in {\cal S}$.

We stress that an asymptotic series is defined by a {\em different limiting 
procedure} than the Taylor one: {\em taking $N$ fixed}, one observes how 
$R_{N}(z)$ behaves for $z \to 0$, $z \in {\cal S}$,  the procedure being 
repeated for all $N \geq 0$ integers. In a Taylor series, however, {\em $z$ is 
fixed} and one observes how the sums $\sum_{n=0}^{N}F_{n}z^{n}$ behave for $N 
\to \infty$. Convergence, a property of the expansion coefficients $F_{n}$, 
may be provable without knowing $F(z)$, to which the series converges. 
However, {\em asymptoticity} can be tested only if one knows {\em both} 
the $F_{n}$ {\em and} $F(z)$.  

The function $F(z)$ may be singular at $z=0$. The coefficients $F_{n}$ in 
(\ref{ptas}) can be defined by
\begin{equation}
F_{n} = \lim_{z \to 0, z \in {\cal S}}   \frac{1}{z^{n}}  \left[F(z) - 
\sum_{k=0}^{n-1}F_{k}z^{k} \right].
\label{coeff2}
\end{equation}
This definition makes sense whenever the asymptotic expansion (\ref{ptas}) 
exists. To define $F_n$, we do without the $n$-th derivative of $F(z)$, 
$z \in {\cal S}$, which may not exist. 

Relation (\ref{ptas}) does {\em not} determine $F(z)$ uniquely; there may be 
many different functions with the same $F_{n}$ coefficients. Note that the 
series (\ref{ptas}) with all the $F_{n}$ vanishing, $F_{n}=0$, is asymptotic to
many functions that are different from the identical zero. Let us denote the 
generic function of this type by $H(z)$; one example is $H(z) = h\, 
{\rm e}^{-c/z}$ with $h \neq 0$ and $c>0$. The expansion with all coefficients 
vanishing is asymptotic to 
$ h\, {\rm e}^{-c/z}$ in the angle $|\arg z| \leq \pi/2-\varepsilon$, where 
$\varepsilon >0$. Then, $F(z)$ and $F(z) + H(z)$ have the same asymptotic 
expansion in the intersection of the two angles, in which the expansions of 
$F(z)$ and $H(z)$ hold.  

The ambiguity of a function given by its asymptotic series is illustrated 
in a more general formulation by the Watson lemma.

%%%%%%%%%%%%%%%%%%%%%%%%%%%%%%%%%%%%%%%%%%%%%%%%%%%%%%%%%%%%%%%%%%%%%%%%%%%
\section{Watson lemma}	\label{sec:watlem}
%%%%%%%%%%%%%%%%%%%%%%%%%%%%%%%%%%%%%%%%%%%%%%%%%%%%%%%%%%%%%%%%%
Consider the following integral 
\begin{equation}
\Phi_{0,c}(\lambda) = \int_{0}^{c} e^{-\lambda
x^{\alpha}}\,x^{\beta -1}f(x) {\rm d}x , 
\label{Laplace}
\end{equation}
where $0<c<\infty$ and  $\alpha > 0, \,\beta > 0 $. Let $f(x) \in 
C^{\infty}{[0,c]}$ and $f^{(k)}(0)$  defined as $\lim_{x \to 0+} f^{(k)}(x)$.
Let $\varepsilon$ be any number from the interval $0 < \varepsilon < \pi/2$. 
 
{\bf Lemma 1} (G.N. Watson): {\em If the above conditions are fulfilled,   
the asymptotic expansion}
\begin{equation}
\Phi_{0,c}(\lambda) \sim \frac{1}{\alpha}\sum_{k=0}^{\infty} 
\lambda^{-\frac{k+\beta}{\alpha}}\, \Gamma
\bigg(\frac{k+\beta}{\alpha}\bigg)\frac{f^{(k)}(0)}{k!} 
\label{Watson}
\end{equation}
 {holds \em for $\lambda \rightarrow \infty, 
\lambda \in S_{\varepsilon}$, where  $S_{\varepsilon}$ is the angle}
\begin{equation}
|\arg \lambda| \leq \frac{\pi}{2} - \varepsilon . 
\label{uhel}
\end{equation}
{\em The expansion (\ref{Watson}) can be differentiated with 
respect to $\lambda$ any number of times.} 

 For the proof see for instance \cite{Jeff}-\cite{Fedo}.
 
{\bf Remark 1:} The perturbation expansion  in powers of $z$ discussed in the 
previous section is obtained by setting $F(z)$ and $1/z$ in the place of 
$\Phi_{0,c}(\lambda)$ and $\lambda$ respectively. The formulae for $F(z)$ 
corresponding to (\ref{Laplace}), (\ref{Watson}) and (\ref{uhel}) can be easily 
found. 

{\bf Remark 2:} The angle (\ref{uhel}) does not depend on 
$\alpha, \,\beta$ or $c$.

{\bf Remark 3:} The factor $\Gamma \bigg(\frac{k+\beta}{\alpha}\bigg)$ makes the 
expansion coefficients in (\ref{Watson}) grow faster with $k$ than those of the 
power series of $f(x)$. 

{\bf Remark 4:}  The expansion coefficients in (\ref{Watson}) are  independent 
of  $c$. This illustrates the impossibility of determining a function from its 
asymptotic expansion, as discussed in the previous section: the same  series is 
obtained for all the integrals along the real axis, having any positive number 
$c$ as the upper limit  of integration.

Below we shall display yet another facet of the above ambiguity, showing  that under plausible assumptions the integration contour in the Laplace-Borel 
 transform can be taken arbitrary in the complex plane. 

%%%%%%%%%%%%%%%%%%%%%%%%%%%%%%%%%%%%%%%%%%%%%%%%%%%%%%%%%%%%%%%%
\section{A modified Watson lemma}	\label{sec:modwat}
%%%%%%%%%%%%%%%%%%%%%%%%%%%%%%%%%%%%%%%%%%%%%%%%%%%%%%%%%%%%
Let  $G(r)$  be a continuous complex function of the form $G(r) = 
r \exp (ig(r))$, where $g(r)$ is a real-valued function given on $0\le r < c$, 
with   $0<c \leq \infty$. Assume that the derivative $G'(r)$ is continuous on 
the interval $0\le r < c$ and a constant $r_0>0$ exists such that 
\begin{equation}
|G'(r)| \le K_1 r^{\gamma_1}, \quad\quad r_0\le r < c,
\label{CK1}
\end{equation}
for a nonnegative $K_1$ and a real $\gamma_1$.

Let the constants $\alpha>0$ and $\beta>0$ be given and assume that the quantities
\begin{equation}
A = \inf_{r_0\leq r < c}\alpha  g(r),\quad\quad \quad B = 
\sup_{r_0 \leq r < c}\alpha g(r)
\label{infsup} 
\end{equation}
satisfy  the inequality
\begin{equation}
B-A < \pi-2\varepsilon,
\label{eps} 
\end{equation}
 where $\varepsilon >0$.

Let the function $f(u)$ be defined along the curve $u=G(r)$ and on the disc 
$|u|<\rho$, where $\rho>r_0$. Assume $f(u)$ to be holomorphic on the disc 
and measurable on the curve. Assume that  
\be
|f(G(r))|\leq K_2 r^{\gamma_2},\quad\quad r_0\le r < c,
\label{CK2}\end{equation}
hold for a nonnegative $K_2$ and a real $\gamma_2$.

Define the function  $\Phi_{b,c}^{(G)}(\lambda)$ for $0 \leq b< c$
by\footnote{This integral exists since we assume that $f(u)$ is measurable along
the curve $u=G(r)$ and bounded by (\ref{CK2}).}  
\begin{equation}
\Phi_{b,c}^{(G)}(\lambda)= \int_{r=b}^{c} e^{-\lambda (G(r))^{\alpha}} 
(G(r))^{\beta-1} f(G(r)) dG(r).
\label{bc}
\end{equation}

{\bf Lemma 2:} {\em If the above assumptions are fulfilled, then 
the asymptotic expansion}
\begin{equation}
\Phi_{0,c}^{(G)}(\lambda) \sim \frac{1}{\alpha} \sum_{k=0}^\infty
\lambda^{-\frac{k+\beta}{\alpha}}\, \Gamma \bigg(\frac{k+\beta}{\alpha} \bigg)
\frac{f^{(k)}(0)}{k!}  
\label{V}
\end{equation}
 {\em holds for $\lambda \rightarrow \infty, \lambda \in \cal T_{\varepsilon}$, where}
\begin{equation}
{\cal T}_\varepsilon = \{\lambda: \lambda=|\lambda| \exp({\rm i} \varphi), \, \, \,  
- \frac{\pi}{2}- A + \varepsilon <\varphi< \frac{\pi}{2} - B- \varepsilon  \}.
\label{calT}
\end{equation}
%%%%%%%%%%%%%%%%%%%%%%%%%%%%%%%%%%%%%%%%%%%%%%%%%%%%%%%%%%%%%%%%%%%%%
\section{Proof of Lemma 2} \label{sec:prooflem}
%%%%%%%%%%%%%%%%%%%%%%%%%%%%%%%%%%%%%%%%%%%%%%%%%%%%%%%%%%%%%%%%%%%%%
\subsection{Proof}\label{subsec:proof}
The conditions stated in section \ref{sec:modwat} assume implicitly that  $c\ge r_0$. We  write:
\be\label{sum}
\Phi_{0,c}^{(G)}(\lambda)=\Phi_{0,r_0}^{(G)}(\lambda)+
\Phi_{r_0,c}^{(G)}(\lambda),
\ee 
and define the new function $\tilde G(r)$ by
\be
\tilde G(r) =\frac{r}{r_0} G(r_0),\quad {\rm for} \,\, 0 \leq r
< r_0; \quad\quad \tilde G(r) = G(r),\quad {\rm for} \,\, r \geq r_0.
\ee
 Since $f(u)$ is holomorphic on the  disc $|u|<\rho$, $\rho>r_0$, Cauchy
 theorem allows us to write
\begin{equation}\label{equal}
\Phi_{0,r_0}^{(G)}(\lambda)=\Phi_{0,r_0}^{(\tilde G)}(\lambda),
\end{equation}
{\em i.e.} the integral along the curved path can be replaced by an integral 
along the straight line $u=r \exp( ig(r_0))$. Futhermore, on the disc  $|u|<r_{0}$, the function $f(u)$ can be expressed in the form
\begin{equation}
f(u) = \sum_{k=0}^N\frac{f^{(k)}(0)}{k!} u^k +r_N(u), \quad  \quad |r_N(u)| \leq C_N |u|^{N+1}.
\label{repf}
\end{equation} 
Then  $\Phi_{0,r_0}^{(\tilde G)}(\lambda)$  can be written as 
\begin{equation}
\Phi_{0,r_0}^{(\tilde G)}(\lambda)=\sum_{k=1}^N I_{0,r_0}^{k}(\lambda)\frac{f^{(k)}(0)}{k!} + \int_{0}^{r_0}\exp{(-\lambda r^\alpha
e^{i \alpha g(r_0)}})\,  e^{i g(r_0)} (re^{i  g(r_0)})^{\beta-1}r_N(\tilde G(r))  dr,
\label{rephi}
\end{equation}
where we defined
\begin{equation}\label{Ibc}
I_{b,c}^{k}(\lambda) = \int_{b}^c (re^{i 
g(r_0)})^{\beta-1+k} \exp{(-\lambda r^\alpha
e^{i \alpha  g(r_0)}})\, e^{i g(r_0)} dr 
\end{equation}
for $0 \leq b<c$.

It is useful to write
\begin{equation}
I_{0,r_0}^{k}(\lambda)=I_{0,\infty}^{k}(\lambda)-I_{r_0,\infty}^{k}(\lambda),
\label{I}
\end{equation}
since the first term, $I_{0,\infty}^{k}(\lambda)$, can be trivially computed. We have
\be\label{Iinf}
I_{0,\infty}^{k}(\lambda) = e^{i g(r_0) (\beta+k)} \int_{0}^\infty r^{\beta+k-1}  \exp{(-\lambda r^\alpha
e^{i \alpha  g(r_0)}})\, dr.
\ee
 From the condition (\ref{eps})  and the definition  (\ref{calT})  it follows that, for $\lambda\in\cal T_{\varepsilon}$,  one has $\mbox{Re}\,[\lambda 
e^{i \alpha  g(r_0)}]>0$.   Therefore we can use the well known  result
\begin{equation}
\int_{0}^{\infty} x^{\delta-1} \exp(-\mu x^{\alpha})\,{\rm d} x = 
\frac{1}{\alpha\,\mu^{\delta/\alpha}}\, 
\Gamma\left(\frac{\delta}{\alpha}\right), 
\label{intGam}
\end{equation}
which  holds for $\mbox{Re} \mu > 0$. Setting $\delta=\beta+k$ and $\mu=\lambda\, e^{i \alpha g(r_0)}$,  we obtain from (\ref{Iinf}) 
\be \label{Ik0inf}
I_{0,\infty}^{k}(\lambda) =\frac{1}{\alpha} \, \lambda^{-\frac{k+\beta}{\alpha}} \, \Gamma\left(\frac{k+\beta}{\alpha}\right).
\ee
By inserting this expression in (\ref{rephi}) and using (\ref{equal}), we write  (\ref{sum}) in the form
\bea\label{dif}
\Phi^{(G)}_{0,c}(\lambda) &\!-\!& \frac{1}{\alpha} \sum_{k=0}^N
\lambda^{-\frac{k+\beta}{\alpha}}\, \Gamma \left(\frac{k+\beta}{\alpha}\right)
\frac{f^{(k)}(0)}{k!}=  \Phi_{r_0,c}^{(G)}(\lambda)-\sum_{k=1}^N I_{r_0, \infty}^{k}(\lambda)
\frac{f^{(k)}(0)}{k!}\nonumber\\  &\!+\!& \int_{0}^{r_0}(re^{i 
g(r_0)})^{\beta-1}r_N(\tilde G(r)) \exp{(-\lambda r^\alpha
e^{i \alpha g(r_0)}})\,  e^{i g(r_0)} dr.
\eea
We now proceed to the estimation of the terms in the right hand side of this relation. 
For $ \Phi_{r_0,c}^{(G)}(\lambda)$  we use the definition (\ref{bc}) and 
notice that for $\lambda \in \cal T_\varepsilon$ the inequality
\be
 \mbox{Re}\, (\lambda r^\alpha e^{i \alpha g(r)}) \geq  |\lambda| r^\alpha \sin\varepsilon
\ee holds. Consequently, we have 
\begin{equation}\label{boundlam}
|e^{-\lambda G(r)^\alpha}|=e^{-{\rm Re}[\lambda G(r)^\alpha]} =
e^{-r^\alpha {\rm Re}[\lambda \,e^{i g(r) \alpha}]}
\leq e^{-|\lambda| r^\alpha \sin \varepsilon}.
\end{equation}
Using also (\ref{CK1}) and (\ref{CK2}),   we obtain 
\begin{equation}
|\Phi^{(G)}_{r_0,c} (\lambda)| \leq K_1 K_2 \int_{r_0} ^\infty
x^{\beta-1+\gamma_1+\gamma_2}e^{-|\lambda| x^\alpha \sin \varepsilon}  dx,
\end{equation}
which, with the transformation $x^\alpha = t$, becomes
\begin{equation}\label{bound2}
|\Phi^{(G)}_{r_0,c}(\lambda)| \leq \frac{K_1 K_2}{\alpha}
\int_{r_0^\alpha}^\infty t^{\frac{\beta-1 +\gamma_1 +\gamma_2
-\alpha +1}{\alpha}} e^{-|\lambda|t \sin\varepsilon} dt . 
\end{equation}
There exists $K^\dagger$ such that 
\begin{equation}\label{boundka}
 t^{\frac{\beta +\gamma_1+\gamma_2-\alpha}{\alpha}} \leq
K^\dagger e^{t \sin \varepsilon}
\end{equation}
on the interval $[r_0,\infty)$. So, Eq. (\ref{bound2}) leads to
\begin{equation}\label{bound1}
|\Phi^{(G)}_{r_0,c} (\lambda)| \leq  \frac{K_1 K_2
K^\dagger}{\alpha (|\lambda|-1) \sin \varepsilon} e^{-(|\lambda|-1)
r_0^\alpha \sin \varepsilon}. 
\end{equation}

The integrals $I_{r_0,\infty}^{k}(\lambda)$ appearing  on the right hand side 
of (\ref{dif}) can be estimated in the same way. By comparing (\ref{Ibc}) with 
(\ref{bc}), it follows that  $I_{r_0,\infty}^{k}(\lambda)$ are obtained from 
$\Phi^{(G)}_{r_0,\infty} (\lambda)$ by the replacements  $G(r)\to 
re^{i g(r_0)}$, $\beta \to \beta+k$ and $f(G(r))\to 1$. Setting  $\beta \to 
\beta+k$, $K_1=K_2 =1$ and $\gamma_1=\gamma_2 =0$, we obtain the bound 
\be
|I_{r_0,\infty}^{k}(\lambda)|\leq \int_{r_0} ^\infty
x^{\beta-1+k}e^{-|\lambda| x^\alpha \sin \varepsilon}  dx\,,
\ee
which can be written, using (\ref{bound2}), as
\be\label{bound3}
|I_{r_0,\infty}^{k}(\lambda)|\leq \frac{
K_k^\dagger}{\alpha (|\lambda|-1) \sin \varepsilon} e^{-(|\lambda|-1)
r_0^\alpha \sin \varepsilon}, 
\ee
for a nonnegative, $k$-dependent, constant $K_k^\dagger$.

Finally, for the last term in the r.h.s. of  (\ref{dif}) we use the bound on 
$r_N$ given in (\ref{repf}) and obtain, by same procedure, the upper bound
\begin{equation}\label{rest}
C_N \int_0^{r_0} x^{\beta-1} x^{N+1} e^{-|\lambda| x^\alpha \sin \varepsilon}dx .
\end{equation}
 This integral can be bounded using (\ref{intGam})  with $\delta=\beta+N+1$ 
 and $\mu = \lambda\sin \varepsilon$, which leads to 
\begin{equation}\label{bound4}
C_N \int_0^{r_0} x^{\beta-1} x^{N+1} e^{-|\lambda| x^\alpha \sin \varepsilon} 
dx= O\left((|\lambda| \sin\varepsilon)^{-\frac{\beta+N+ 1}{\alpha}}\right). 
\end{equation}
The estimates (\ref{bound1}),   (\ref{bound3}) and  (\ref{bound4}), inserted in the r.h.s. of (\ref{dif}), show that
\begin{equation}\label{difest}
\Phi_{0,c}^{(G)}(\lambda) - \frac{1}{\alpha} \sum_{k=0}^N
\lambda^{-\frac{k+\beta}{\alpha}} \Gamma \bigg(\frac{k+\beta}{\alpha}\bigg)
\frac{f^{(k)}(0)}{k!}=O\bigg(|\lambda|^{-\frac{\beta+N+1}{\alpha}}\bigg)
\end{equation}
for $\lambda \rightarrow \infty,\, \lambda \in \cal T_\varepsilon$. This 
completes the proof of (\ref{V}).
%%%%%%%%%%%%%%%%%%%%%%%%%%%%%%%%%%%%%%%%%%%%%%%%%%%%%%%%%%%%%%%%%%%%%%%%%%%%%%%%%%%%%%%%
\subsection{Optimality}
There is a question whether the angle $\cal T_\varepsilon$ given in Lemma 2
can be enlarged. We show that the angle is maximal by proving that outside 
$\cal T_\varepsilon$ the relation (\ref{difest}) does not hold. 
\par Let us take a special case $g(r)=0$ on  $[0,\rho]$, where $0<\rho < c < 
\infty$, $g(c)=\pi/4$. Let $0 \leq g(r) \leq \pi/4$ be fulfilled for $r \in 
[\rho,c]$, $g(r)$ being a smooth function. We choose a special function $f$ by 
taking $f(u)=f_1$ for $|u| <\rho$ and $f(G(r))=f_2$ for $r \in [ \rho,c]$ where 
$f_1,f_2$ are two different, nonzero constants. We take $\alpha=\beta =1$. 
Certainly we have $A= 0, B = \pi/4$. The assumptions of Lemma 2 are fulfilled 
and we obtain
\begin{equation} \label{angle1}
{\cal T}_\varepsilon = \{ \lambda: \lambda= |\lambda| \exp({\rm i}  
\varphi),-\frac{\pi}{2}+\varepsilon < \varphi< \frac{\pi}{4} - \varepsilon\} \, , 
\end{equation}
where $\varepsilon$ is an arbitrary positive number.   
Define, as in the Lemma 2, 
\begin{equation}
\Phi_{0,c}^{(G)}(\lambda)= \int_0^c\, e^{-\lambda G(r)}\,
f(G(r))\, dG(r). 
\end{equation}
Now we choose a ray that lies outside $\cal T_\varepsilon$ , 
\begin{equation}
 L_1= \{ \lambda:\lambda=|\lambda| \exp(-i (\pi/2+\delta)) \}
\end{equation}
where $0 < \delta < \pi/4 $. We shall show that the function
$\Phi_{0,c}^{(G)}(\lambda)$ is unbounded along the ray $L_1$.
\par This function can be written
\begin{equation}
\Phi_{0,c}^{(G)}(\lambda)=[f_1+(f_2-f_1)\exp(-\lambda \rho)]/\lambda -
f_2 \exp[- \lambda c \exp(i g(c))]/\lambda.  \label{angle4} 
\end{equation}
Now for $\lambda \in L_1$ we have
\begin{equation}
|\exp(- \lambda \rho)/\lambda|=\exp( -|\lambda| \rho |
\cos(-\pi/2-\delta)) /|\lambda|, 
\label{5}
\end{equation}
which is divergent for $|\lambda| \rightarrow \infty$.
The last term of (\ref{angle4}), 
\begin{equation}
|f_2/\lambda| | \exp[- \lambda c \exp(i g(c))]|=
|f_2/\lambda|\exp[-|\lambda| c \cos(-\pi/4-\delta)]
\label{angle7}
\end{equation}
converges to zero as $|\lambda| \rightarrow \infty$.
It follows that Lemma 2 does not apply for $\lambda \in L_1,
|\lambda | \rightarrow \infty$ (see (\ref{V})).
\par Now we choose another ray, $L_2$, which also lies outside 
$\cal T_\varepsilon$:
\begin{equation}
 L_2= \{ \lambda:\lambda=|\lambda| \exp(i (\pi/4+\delta)) \} . 
\label{angle8}
\end{equation}
Certainly
\begin{equation}
|\exp(- \lambda \rho)|=\exp( -|\lambda| \rho
\cos(\pi/4+\delta)),
\end{equation}
which converges to zero for $\lambda \in L_2, |\lambda|
\rightarrow \infty. $ For the last term we have
\begin{equation}
|f_2/\lambda||\exp[- \lambda \, c \exp(i g(c))]|=
|f_2/\lambda|\exp[-|\lambda| \, c \cos(\pi/2 +\delta)] ,
\end{equation}
which is divergent for $\lambda \in L_2, |\lambda| \rightarrow \infty.$
%%%%%%%%%%%%%%%%%%%%%%%%%%%%%%%%%%%%%%%%%%%%%%%%%%%%%%%%%%%%%%%%%%%%%%%%%%%%%
\section{Remarks on Lemma 2 and its proof}	\label{sec:remlem}
{\bf Remark 5:}  Up to now we have assumed that $c>\rho$, where $\rho$ is the 
radius of the convergence disc of $f(u)$. If $c<\rho$, then the entire contour 
from 0 to $c$ can be deformed up to a straight line, and we have instead of 
(\ref{dif}):
\bea\label{dif1}
\Phi_{0,c}^{(G)}(\lambda) &\!-\!& \frac{1}{\alpha} \sum_{k=0}^N
\lambda^{-\frac{k+\beta}{\alpha}}\, \Gamma \left(\frac{k+\beta}{\alpha}\right)
\frac{f^{(k)}(0)}{k!}= -\sum_{k=1}^N I_{c, \infty}^{k}(\lambda)\frac{f^{(k)}(0)}{k!}\nonumber\\  &\!+\!& \int_{0}^{c}(re^{i 
g(c)})^{\beta-1}r_N(\tilde G(r)) \exp{(-\lambda r^\alpha
e^{i \alpha g(c)}})\,  e^{i g(c)} dr.
\eea
The integrals $I_{c, \infty}^{k}(\lambda)$ and the last term in (\ref{dif1}) 
can be estimated  as above, with the replacement of $r_0$ by $c$ in 
(\ref{bound3}) and (\ref{rest}). These parameters do not appear in the 
asymptotic expansion in the l.h.s. of  (\ref{difest}), which preserves its 
form. The difference is that the exponential suppression of the remainder 
depends now only on $c$. The reason is that we can choose $r_{0} = c$, and
the conditions (\ref{CK1}) - (\ref{CK2}) are empty. 

{\bf Remark 6:} Watson's lemma is obtained in the special case where the 
integration contour becomes a segment of the real positive semiaxis, {\em 
i.e.} $g(r)\equiv 0$, and $f(r) \in C^\infty[0,c]$. In fact it is enough 
to assume that $f(r) \in C^\infty[0,c']$ with $c'<c$, where $f(r)$ is 
measurable for $r \in (c', c)$ and bounded according to (\ref{CK2}).  

\begin{figure}\begin{center}\vspace{0.2cm}
\includegraphics[width=7.5cm]{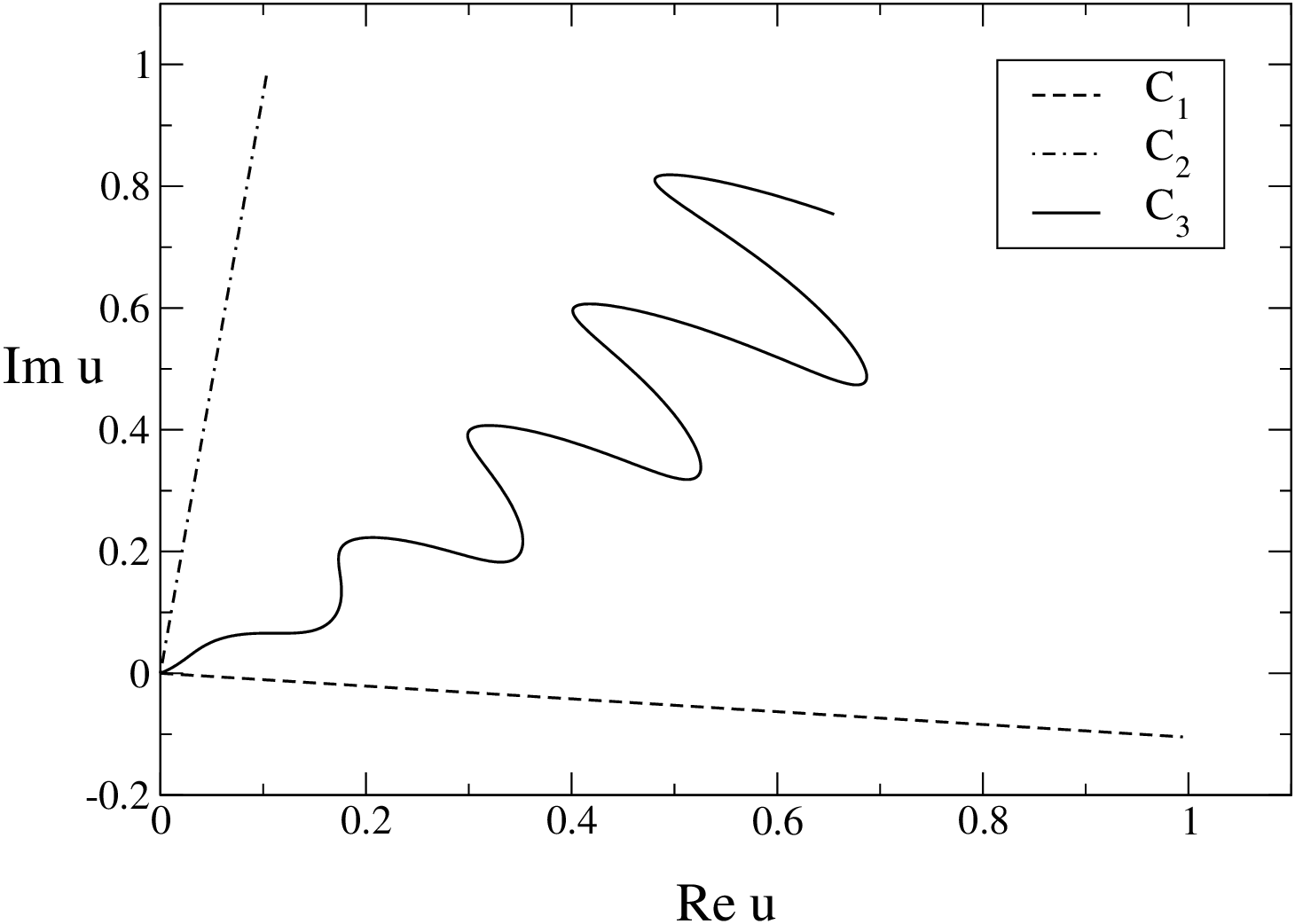}
\caption{\label{fig:Curves} Three examples of integration contours 
in the $u$-plane.} \end{center}\end{figure}

\begin{figure}[thb]\begin{center}\vspace{0.2cm}
\includegraphics[width=6.5cm]{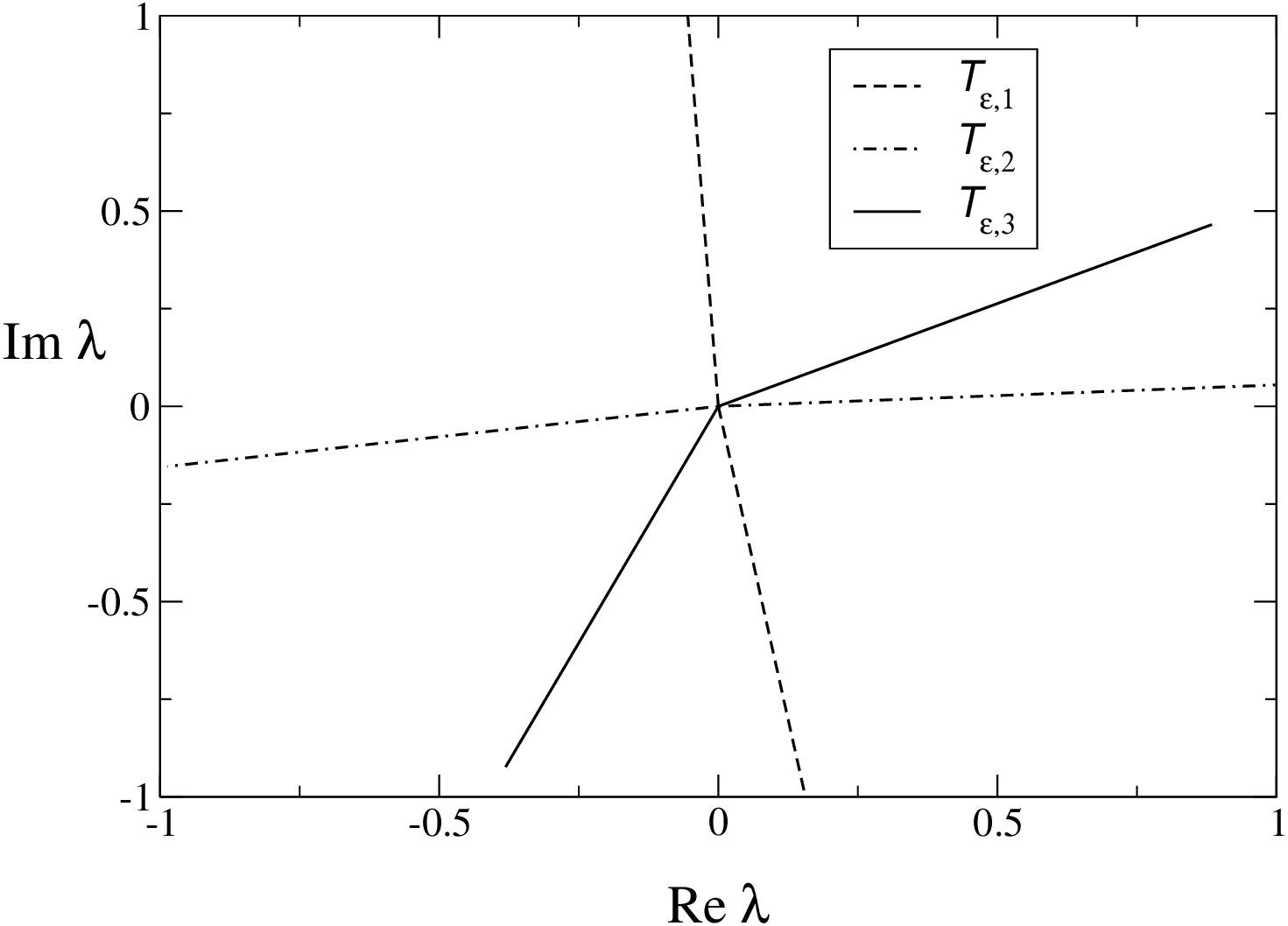}\hspace{1cm}\includegraphics[width=6.5cm]{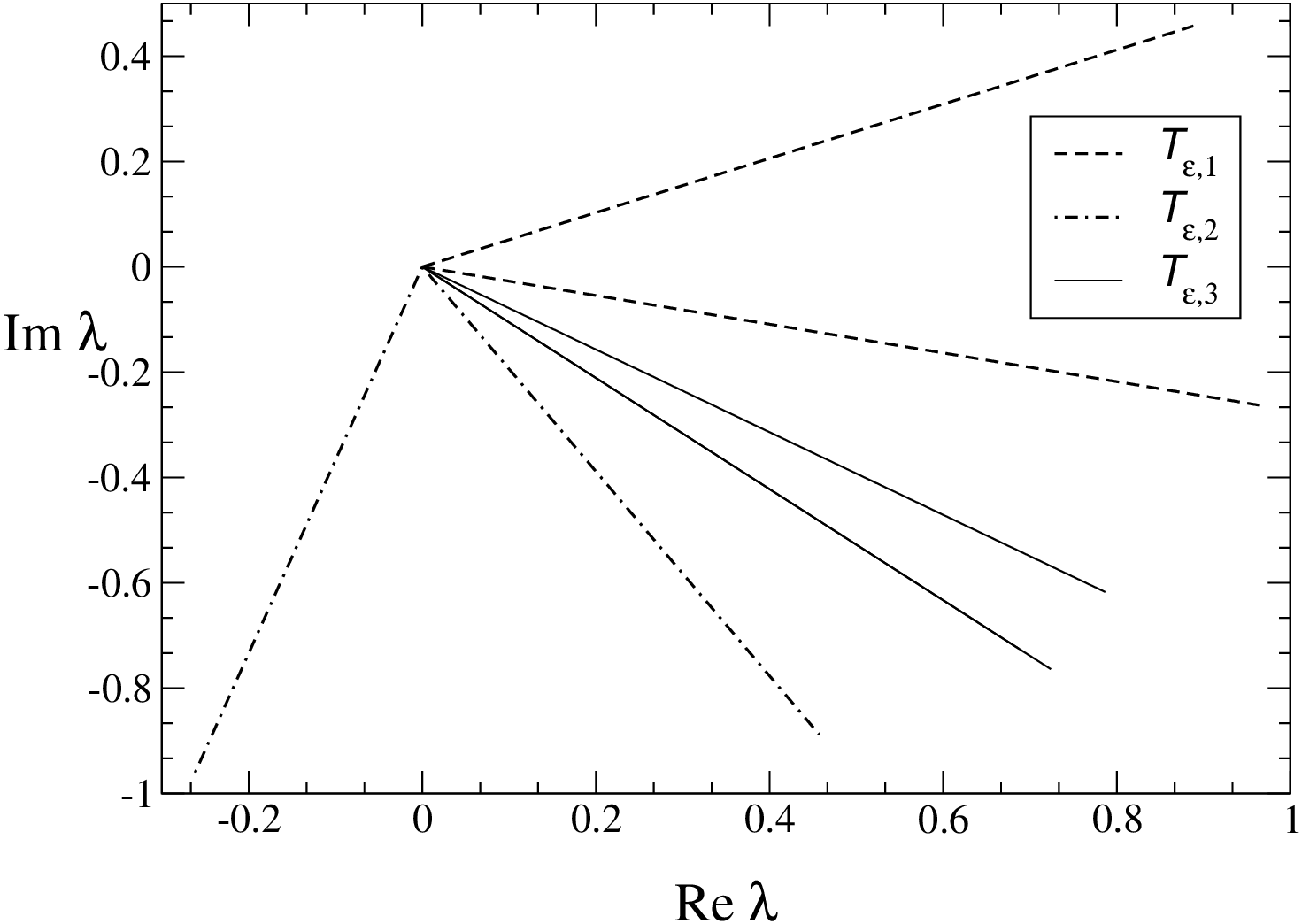}
\caption{\label{fig:Sectors}  Three regions ${\cal T}_{\varepsilon, 
k}$ in the complex $\lambda$ plane given by Eq. (\ref{calT}), for the three contours 
${\cal C}_k$ shown in Fig. \ref{fig:Curves}. If $\varepsilon$ is small, 0.05 say (left 
panel),  all the three angles are large, though ${\cal T}_{\varepsilon, 3}$ is smaller 
than the other two. For $\varepsilon=1.2$ (right panel), $\varepsilon$ is near the 
upper limit 1.27 imposed by condition (\ref{eps}) for $k=3$; as a consequence, the 
angle ${\cal T}_{\varepsilon, 3}$ is very narrow.}\end{center}\end{figure}

{\bf Remark 7:} For applications in perturbation theory, we set $\lambda=1/z$  
in (\ref{bc}). For simplicity, we take the particular case $\alpha=\beta=1$.  
Lemma 2 then implies that  the function
\begin{equation}
F^{(G)}_{0,c}(z)= 
\int_{r=0}^c e^{-G(r)/z}\, f(G(r))\, dG(r) 
\label{abz11}
\end{equation}
has the asymptotic expansion
\begin{equation}
F_{0,c}^{(G)}(z) \sim \sum_{k=0}^\infty z^{k+1} f^{(k)}(0)  
\label{asyF} 
\end{equation}
 for $z \rightarrow 0$ and $z \in \cal Z_\varepsilon$, where 
\begin{equation}
{\cal Z}_\varepsilon =\{z: z=|z| \exp{(i \chi)},\, -\frac{\pi}{2}+B + \varepsilon <\chi< 
\frac{\pi}{2}+A - \varepsilon \}.
\label{calZ}
\end{equation}
%%%%%%%%%%%%%%%%%%%%%%%%%%%%%%%%%%%%%%%%%%%%%%%%%%%%%%%%%%%%%%%%%%%%%%%%%%%%%%%%
{\bf Remark 8:} The parameter $\varepsilon$ in the condition (\ref{eps}) is 
limited by $0< \varepsilon<\pi/2-(B-A)/2$, but is otherwise rather arbitrary. 
Note however that the upper limit of $\varepsilon$ may be considerably less 
than $\pi/2$, being dependent on the value of $B-A$. This happens, in 
particular, if the integration contour is bent or meandering. We illustrate 
this in Fig. \ref{fig:Curves} by three contours ${\cal C}_k$, $k=1, 2,3$ given 
by the parametrizations $u=G_k(r)=r \exp{(ig_k(r))}$  with $g_1(r)=-\pi/30$, 
$g_2(r)=\pi/2-\pi/30$, and $g_3(r)=0.6+0.25 r +0.2 \sin 9\pi r$. We consider 
$\alpha=\beta=1$ and take the upper limit $c=1$. Then we have $B-A=0$ for the 
constant functions $g_1$ and $g_2$, where $\varepsilon$ can be any number in 
the range $(0,\,\pi/2)$, while for $g_3$ the difference $B-A$ equals 0.59, and 
the condition (\ref{eps}) gives $0< \varepsilon< 1.27$. As follows from 
(\ref{calT}) and can  be seen in  Fig. \ref{fig:Sectors} left, if 
$\varepsilon$ is near zero, the domains ${\cal T}_{\varepsilon, k}$,
 $k=1,2,3$, are large. However, Eqs. (\ref{bound1}) and (\ref{bound3}) show 
that the bounds on the remainder of the truncated series  are loose in this 
case. If on the other hand $\varepsilon$ is near its upper limit (see Fig. 
\ref{fig:Sectors} right), the bounds are tight, but this has a cost in the 
fact that the ${\cal T}_{\varepsilon, k}$, $k=1,2,3,$ the angles of validity 
of the expansion, are small. 

%%%%%%%%%%%%%%%%%%%%%%%%%%%%%%%%%%%%%%%%%%%%%%%%%%%%%%%%%%%%%%%%%%%%%%%%%%%%%%%

%%%%%%%%%%%%%%%%%%%%%%%%%%%%%%%%%%%%%%%%%%%%%%%%%%%%%%%%%%%%%%%%%%%%%%%%%%%%%%
%%%%%%%%%%%%%%%%%%%%%%%%%%%%%%%%%%%%%%%%%%%%%%%%%%%%%%%%%%%%%%%%%%%%%%%%%%%%%%
{\bf Remark 9:}  We note that  the parametrization $G(r)=r \exp{(ig(r))}$ does not include contours that cross 
a circle centred at $r=0$ either touching or doubly intersecting it, so that the derivative $G'(r)$ does not exist or is not bounded. In particular, this 
parametrization does not include the contours 

(i) that, starting from the origin and reaching a value $r_1$ of $r$, return 
back to a certain value $r_2<r_1$, closer to the origin, and  
 
(ii) whose one or several parts coincide with a part of a circle centred at 
the origin. 

The contours under (i) can be included by representing the integral as a sum of 
several integrals along individual contours that separately fulfill the 
conditions of Lemma 2. In particular, if the integration contour returns back 
to the origin, the expansion coefficients all vanish. The cases under item 
(ii) can be treated by choosing an alternative  parametrization of the curve.

{\bf Remark 10:} Extensions of the Watson lemma to the complex plane  were
considered briefly  by H. Jeffreys in Ref. \cite{Jeff}, under somewhat 
different conditions. 

{\bf a)} Identical is the  condition  requiring that the function 
$f(u)$ is holomorphic on a disc, $|u|<R$. 

{\bf b)} In Ref. \cite{Jeff}, the integration is performed along a right angle 
in the complex $u$-plane, first along the real axis and then along a segment 
parallel to the imaginary axis, up to the upper point of the integration 
contour. So, it is assumed that the integral along this right angle exists, in 
particular, that the function $f(u)$ is defined along it. In fact, $f$ is
assumed to be analytic in the region bounded by the initial contour and this
right angle, except for some isolated singularities. The condition 
of Lemma 2 requires only that, outside the circle, $f(u)$ is defined along the 
curve of the original contour of integration. These weaker assumptions imply  
that our approach is more general.

{\bf c)} Lemma 2 and its proof allow us to use the bounds (\ref{bound1}), 
(\ref{bound3}), and (\ref{bound4}) to obtain an estimate for the difference 
between $\Phi^{(G)}_{0,c}(\lambda)$ and its asymptotic expansion (\ref{V}) in 
terms of the constants introduced in  the conditions (\ref{CK1}), (\ref{infsup}) 
and (\ref{CK2}). 

{\bf d)} Finally, we remind the reader that the proof of Lemma 2 in subsection 
\ref{subsec:proof} allowed us to obtain a remarkable correlation between the 
strength of the bounds on the remainder and the size of the angles where the 
asymptotic expansion is valid. Indeed,  (\ref{bound1}), (\ref{bound3}), and 
(\ref{bound4}) depend on the parameter $\varepsilon$, which  determines the 
angles  $\cal T_\varepsilon$ and $\cal Z_\varepsilon$, see (\ref{calT}) and 
(\ref{calZ}) respectively. As was pointed out in Remark 8, the larger the 
angle of validity, the looser the bound, and vice versa.  
%%%%%%%%%%%%%%%%%%%%%%%%%%%%%%%%%%%%%%%%%%%%%%%%%%%%%%%%%%%%%%%%%
 \section{Remarks on perturbative QCD} \label{sec:remqcd}
%%%%%%%%%%%%%%%%%%%%%%%%%%%%%%%%%%%%%%%%%%%%%%%%%%%%%%%%%%%%%%%%%
We take the  Adler function \cite{Adler} 
\begin{equation}
{\cal D}(s) =- s\frac{{\rm d}\Pi(s)}{{\rm d}s}-1\,,
\label{calD}
\end{equation}
to discuss applications of Lemma 2, where $\Pi(s)$ is the polarization 
amplitude defined from
\begin{equation} 
i \int{\rm d}^{4}x e^{iq.x}<0|T\{V_{\mu}(x)V_{\nu}(0)^{\dag}\}|0>
= (q_{\mu}q_{\nu}-g_{\mu\nu}q^{2})\Pi(s).
\label{correl}
\end{equation}
Here $s=q^2$ and $V_\mu$ is the vector current for light ($u$ or $d$) quarks. 
 
In accordance with general principles \cite{BogSh, Adler}, ${\cal D}(s)$ is 
real analytic in the complex $s$-plane, except for a cut along the time-like 
axis produced by  unitarity. In perturbative QCD, any finite-order aproximant 
has cuts along the time-like axis, while the renormalization-group
improved expansion,
\be\label{Dpert}
{\cal D}(s) = D_1 \,\alpha_s(s)/\pi +  D_2 \,(\alpha_s(s)/\pi)^2 + 
D_3 \,(\alpha_s(s)/\pi)^3  + \ldots \,,  
\ee 
has in addition  an unphysical singularity  
due to the Landau pole in the running coupling $\alpha_s(s)$. 
According to present knowledge, (\ref{Dpert}) is 
divergent,  $D_n$ growing as $n!$ at large $n$ 
\cite{Mueller1992}-\cite{BenekePR}. 

%%%%%%%%%%%%%%%%%%%%%%%%%%%%%%%%%%%%%%%%%%%%%%%%%%%%%%%%%%%%%%%%%%%%%%%%%%%%
\subsection{Ambiguity of the perturbative QCD}
To discuss the implications of Lemma 2, we can define the  Borel transform 
$B(u)$  by  \cite{Neubert1996}
\be\label{B}
B(u)= \sum\limits_{n\ge 0} b_n \,u^n,\quad\quad \quad 
b_n=\frac{D_{n+1}}{\beta_0^n\,n!}\,.
\ee
It is usually assumed the series (\ref{B}) is convergent on a disc of
nonvanishing radius (this result was rigorously proved by David et al. 
\cite{David} for the scalar $\varphi^4$ theory).
This is exactly what is required in Lemma 2 for the Borel transform.

If we adopt the assumption that the series (\ref{Dpert}) is asymptotic, Lemma 2 
implies a large freedom in recovering the true function from its perturbative 
coefficients. Indeed, taking for simplicity $\alpha=\beta=1$ in (\ref{bc}), we 
infer that all the  functions ${\cal D}^G_{0,c}(s)$ of the form
\be\label{BDG}
{\cal D}^G_{0,c}(s)=\frac{1}{\beta_0} \int_{r=0}^c e^{-\frac{G(r)}{\beta_0\,
a(s)}}\, B(G(r))\, {\rm d}G(r) \, ,
\ee  
where $a(s)=\alpha_s(s)/\pi$,  admit the asymptotic expansion
\be\label{DGpert}
{\cal D}^G_{0,c}(s) \sim \sum\limits_{n=1}^{\infty} D_n \, (a(s))^n, \quad 
\quad \quad  a_{s}(s) \to 0, 
\ee
in a certain domain of the $s$-plane, which follows from (\ref{calT}) and the 
expression of the running coupling $a(s)$ given by the renormalization group.

As mentioned above,  Lemma 2 imposes weak conditions on $B(u)$ and on the 
integration contour. 
 Outside the convergence disc of (\ref{B}), the form of $B(u)$ (denoted $f(u)$ in section
\ref{sec:modwat}) is largely arbitrary, being restricted only by the rather
 weak conditions of Lemma 2. 
If the function $B(u)$ defined by (\ref{B}) admits an analytic
continuation outside the disc (which is not necessary for Lemma 2 to apply), 
the analytic continuation can be used as input in the integral
representation  (\ref{DGpert}). Then, if the 
contour passes through the analyticity domain, ${\cal B}$ say, more specific
properties  of ${\cal D}^G_{0,c}(s)$ in the coupling plane can be derived, in 
analogy with the case of Borel summable functions (see \cite{Sokal}).

The integral (\ref{BDG}) reveals the large  ambiguity  of the resummation 
procedures having the same asymptotic expansion in perturbative QCD: no 
particular function of the form ${\cal D}^G_{0,c}(s)$  can be a  priori 
preferred when looking for the true  Adler function. 

The proof of Lemma 2 shows that the form and length of the contour, as well as
the values of $B(u)$ outside the convergence disc, do not 
affect the series (\ref{DGpert}), contributing only to the exponentially 
suppressed remainder. As seen from the r.h.s. of (\ref{dif}) or (\ref{dif1}), 
the terms to be added to (\ref{DGpert}) are $\Phi^G_{r_0,c} (\lambda)$ and 
$I^k_{r_0, \infty}(\lambda)$, where $\lambda = 1/(\beta_{0}a(s))$. The 
estimates in (\ref{bound1}) and (\ref{bound3}) imply the remainder to 
(\ref{DGpert}) to have the form  $h\, {\rm exp}(- d/\beta_{0} a(s)) \sim h 
\left(-\Lambda^{2}/s \right)^{d}$, where we used the running coupling to one 
loop.  The quantities $h$ and $d>0$  depend on 
the integration contour and on the values of $B(u)$ outside the disc, which 
can be chosen rather freely. So, 
(\ref{BDG}) contains arbitrary power terms, to be added to 
(\ref{DGpert}).\footnote{The connection between power corrections and 
Borel-Laplace integrals on a finite range was discussed also in \cite{Chyla}.}  
%%%%%%%%%%%%%%%%%%%%%%%%%%%%%%%%%%%%%%%%%%%%%%%%%%%%%%%%%%%%%%%%%%%%%%%%%%%%
\subsection{Optimal conformal mapping and analyticity}

In problems of divergence and ambiguity, the location of singularities of 
${\cal D}(s)$ and $B(u)$ in the $a(s)$-plane and, respectively, in the 
$u$-plane, is of importance.

 Some information 
about the singularities of ${B(u)}$ is obtained from certain classes of Feynman 
diagrams, which can be explicitly summed \cite{Beneke1993}-\cite{BenekePR}, and 
from general arguments based on renormalization theory 
\cite{Mueller1992,BeBrKi}. This analysis shows that $B(u)$  has branch points 
along the rays $u \geq 2$ and $u\leq -1$ (IR and UV renormalons respectively). 
Other (though nonperturbative) singularities, for $u \ge 4$, are produced by 
instanton-antiinstanton pairs. Due to the singularities at $u>0$, the series 
(\ref{Dpert}) is not Borel summable. 
 Except the above mentioned singularities of $B(u)$ 
on the real axis of the $u$-plane, however, no other singularities are known; 
it is usually assumed that, elsewhere, $B(u)$ is holomorphic. 

To treat the analyticity properties of $B(u)$, the method of optimal conformal 
mapping \cite{SorJa} is very useful. If the analyticity domain ${\cal B}$ is larger than the disc of 
convergence of (\ref{B}), one replaces (\ref{B}) by the expansion 
\be\label{Bw}
B(u)=\sum_{n\ge 0} c_n \,w^n,
\ee
where $w=w(u)$ (with $w(0)=0$) maps ${\cal B}$ (or a part of it) onto the disc $|w|<1$, on which 
(\ref{Bw}) converges. (\ref{Bw}) has better convergence properties than 
(\ref{B}): in \cite{SorJa}, Schwarz lemma was used to prove that the larger 
the region mapped by $w(u)$ onto $|w|<1$, the faster the convergence rate of 
(\ref{Bw}). 

If $w(u)$ maps the whole  ${\cal B}$ onto the disc $|w|<1$, (\ref{Bw}) converges on 
the whole region ${\cal B}$ and, as shown in \cite{SorJa}, its convergence rate 
is the fastest.\footnote{This mapping is called optimal. In the particular case 
when ${\cal B}$ is the $u$-plane cut along the rays $u<-1$ and $u>2$, the 
optimal mapping reads $w(u)=(\sqrt{1+u}-\sqrt{1-u/2})/(\sqrt{1+u}+\sqrt{1-u/2})$ 
\cite{SorJa,CaFi}. Note that the expansion (\ref{Bw}) takes into account only 
the {\em location} of the singularities of $B(u)$. Ways of accounting for their 
{\em nature} can be found in \cite{Soper, CaFi}.} Then, the region of convergence 
of (\ref{Bw}) coincides with the region ${\cal B}$ of analyticity of $B(u)$.

Optimal conformal mapping allows one to express analyticity in terms of
convergence.  
 Inserting (\ref{Bw}) into (\ref{BDG}) we have 
\be\label{BDGw}
{\cal D}^G_{0,c}(s)=\frac{1}{\beta_0} \int_{r=0}^c e^{-\frac{G(r)}{\beta_0\,
a(s)}}\, \sum_{n\ge 0} c_n \, [w(G(r))]^n \, {\rm d}G(r) \,.
\ee  
This expression admits the same asymptotic expansion (\ref{DGpert}). However, 
containing powers of the variable  $w$, it implements more information about 
the  singularities of the true Borel transform $B(u)$ than the series (\ref{B}) 
in powers of $u$, even at  finite orders. So, one expects that the finite-order 
approximants of (\ref{BDGw}) will provide a  more accurate description of 
the physical function searched for \cite{CaFi, CvLe}. 

\subsection{Piecewise analytic summation}
The choice of the integration contour may have a fateful impact on analyticity. 
In \cite{HoMa,BrMa}, two different contours in the $u$-plane are chosen for the summation of some class of diagrams (the so-called renormalon chains \cite{Beneke1993, Broad1993}): one contour,  parallel and close to the positive semiaxis, is adopted for $a(s)>0$,  another one,  parallel and close to the negative semiaxis, is taken when
$a(s)<0$. As was expected, and later proved in \cite{CaFi4}, analyticity is lost 
with this choice, the summation being only piecewise analytic in $s$. Although 
this summation represents only a part of the full correlator, it is preferable 
to approximate an analytic correlator by a function which is also analytic, 
since analyticity in the $s$-plane is needed for relating perturbative QCD in 
the Euclidean region to measurable quantities on the time-like axis. 

On the other hand, as shown in \cite{CaNe, CaFi3}, the Borel summation with the Principal Value (PV) prescription of  
the same class of diagrams  admits an 
analytic continuation to the whole $s$-plane, being consistent with analyticity 
except for an unphysical cut along a segment of the space-like axis, related to the Landau pole.  In this sense, PV is an appropriate prescription.

%%%%%%%%%%%%%%%%%%%%%%%%%%%%%%%%%%%%%%%%%%%%%%%%%%%%%%%%%%%%%%%%%%
\section{Concluding remarks} \label{sec:conrem} 
%%%%%%%%%%%%%%%%%%%%%%%%%%%%%%%%%%%%%%%%%%%%%%%
The main result of our work is a modified form of Watson's lemma on 
asymptotic series. This result,  referred to as Lemma 2, holds 
if the function  $f(u)$ (which corresponds to  the Borel transform of a 
 QCD correlator) is analytic in a disc  and is defined along a contour, on which
  it  satisfies rather weak conditions, which are specified in section 
  \ref{sec:modwat}. 

 Our result  emphasizes the great 
ambiguity of the summation prescriptions that are allowed if the perturbation 
expansion in QCD is regarded as asymptotic.  The contour  of the   integral 
representing  the QCD correlator and the function $B(u)$  can be chosen very 
freely outside the convergence disc. 

We kept our discussion on a general level, bearing in mind that little is known, 
in a rigorous framework, about the 
analytic properties of the QCD correlators in the Borel plane. If some specific 
properties are known or assumed, the integral representations discussed in ths paper
will have additional analytic properties, known to be shared also by the 
physical amplitudes.

Lemma 2 proved in this paper may also be useful in other branches of physics 
where perturbation series are divergent. 

\begin{acknowledgments} 
We thank Pavel Kol\'{a}\v{r} for useful discussions and comments on the manuscript. 
One of us (I.C.) thanks Prof. Ji\v{r}\'i Ch\'yla and the Institute of Physics of the 
Czech Academy in Prague for hospitality. One of us (J.F.) thanks Prof. Piotr
R\c aczka and the Institute of Theoretical Physics of the Warsaw University for
hospitality. This work was supported by CNCSIS in the frame of the Program Idei, Contract Nr. 464/2009, and by the Projects No. LA08015 of the Ministry of 
Education and AV0-Z10100502 of the Academy of Sciences of the Czech Republic.
\end{acknowledgments}

%%%%%%%%%%%%%%%%%%%%%%%%%%%%%%%%%%%%%%%%%%%%%%%%%%%%%%%%%%%

\end{document}